\documentclass[sigconf]{acmart}

\usepackage{booktabs} 

\usepackage{comment}
\setcopyright{rightsretained}
\usepackage[caption=false]{subfig}
\usepackage{amsmath}
\usepackage{comment}
\acmDOI{xx.xxx/xxx_x}

\acmISBN{978-1-4503-5933-7/19/04}

\acmConference[SAC'19]{ACM SAC Conference}{April 8-12, 2019}{Limassol, Cyprus} 
\acmYear{2019}
\copyrightyear{2019}

\acmArticle{4}
\acmPrice{15.00}

\begin{document}
\title{Using Machine Learning for Handover Optimization in Vehicular Fog Computing}

\author{Salman Memon}
\affiliation{%
  \institution{McGill University}
  \city{Montreal} 
  \state{Quebec} 
}
\email{salman.memon@mail.mcgill.ca}

\author{Muthucumaru Maheswaran}
\affiliation{%
  \institution{McGill University}
  \city{Montreal} 
  \state{Quebec} 
}
\email{maheswar@cs.mcgill.com}

\begin{abstract}
Smart mobility management would be an important prerequisite for future fog computing systems. In this research, we propose a learning-based handover optimization for the Internet of Vehicles that would assist the smooth transition of device connections and offloaded tasks between fog nodes. To accomplish this, we make use of machine learning algorithms to learn from vehicle interactions with fog nodes. Our approach uses a three-layer feed-forward neural network to predict the correct fog node at a given location and time with 99.2 \% accuracy on a test set. We also implement a dual stacked recurrent neural network (RNN) with long short-term memory (LSTM) cells capable of learning the latency, or cost, associated with these service requests. We create a simulation in JAMScript using a dataset of real-world vehicle movements to create a dataset to train these networks.  We further propose the use of this predictive system in a smarter request routing mechanism to minimize the service interruption during handovers between fog nodes and to anticipate areas of low coverage through a series of experiments and test the models' performance on a test set. 
\end{abstract}

%
%


\copyrightyear{2019} 
\acmYear{2019} 
\setcopyright{acmcopyright}
\acmConference[SAC '19]{The 34th ACM/SIGAPP Symposium on Applied Computing}{April 8--12, 2019}{Limassol, Cyprus}
\acmBooktitle{The 34th ACM/SIGAPP Symposium on Applied Computing (SAC '19), April 8--12, 2019, Limassol, Cyprus}
\acmPrice{15.00}
\acmDOI{10.1145/3297280.3297300}
\acmISBN{978-1-4503-5933-7/19/04}

\maketitle

\section{Introduction}

Catering to the computationally intensive, data-driven and latency-sensitive applications in smart cities and the Internet of Things (IoT) is expected to be a considerable challenge for the existing cloud computing model \cite{shi2016edge}. Fog computing provides a viable solution to these problems by placing compute and storage resources closer to the edge of the Internet. This provides cloud style on-demand processing for a myriad of applications such as real-time guidance for autonomous cars, video surveillance and real-time analytics \cite{varshney2017demystifying}.

One of the important beneficiaries of fog computing would be the Internet of Vehicles \cite{gerla2014internet}. Fog servers installed as roadside units (RSUs) would provide a shared computing backend that is accessible to the vehicles with minimal access latency and available to deploy a variety of different applications. However, because the fog servers are static and vehicles mobile, we have an association problem to determine which vehicles are connected to which fogs. The criteria for association could depend on how computation and data are split across the fogs. 

Consider an example scenario where a fog server placed in a road segment would only be responsible for that portion of the road. This would force the vehicles to move from one fog to another as they move. Due to load fluctuations, the best fog at a given time and location would vary.

In this paper, we present a machine learning based fog location and cost predictor to help optimize the transition of connected vehicles between fog nodes. A fog location predictor can be used by a vehicle to determine the best fog server it should be connecting to at a given time and location. The cost predictor tells the vehicle the cost of using a particular fog at a given time and location. We use the latency of servicing a request as the cost.

Using the predicted fog location and cost, we can proactively select the serving fogs such that the disruptions during fog-to-fog transitions are minimized. To accomplish this, we can use a variety of strategies including sending a duplicate request to the fog close to the switchover point and duplicating the data to the next fog. A lot of the enabling mechanisms for proactively using the fogs are supported by JAMScript,  a programming language, and middleware for edge-oriented IoT systems \cite{wenger2016programming}.

The dataset for training and testing the machine learning models were created using traces provided from Shanghai Jiao Tong University \cite{dataset}. A vehicular fog simulator developed in JAMScript was used to add fog access and cost information to vehicles moving according to these traces.

We make use of a two-layer feed-forward neural network predict the best fog node and a dual stacked Reccurent Neural Network (RNN) with Long Short-Term Memory (LSTM) cells to forecast the expected performance, or cost, at a location and time. These learning algorithms find the optimal fog at any location and time and are trained at the cloud. The trained models can be downloaded to the vehicles, thereby allowing them to the predictions while making fog selection for remote tasks independently. 

In the following section, we present an overview of the existing approaches proposed for mobility management in a fog computing scenario. In section \ref{architecture}, we describe the system architecture and how the components of our proposed solution would interact with a governing middleware. Section \ref{sec:fogandcost} describes the data modeling methods and section \ref{sec:requestrouting} presents a request routing method that takes advantage of them to anticipate the next fog node a device will interact with. Section \ref{sec:implementation} discusses the implementation steps and section \ref{sec:exp} shows a variety of experiments to gauge the performance of our learning models.   

\section{Related Work and Our Contribution}
Mobility management for fog computing is a topic that has received considerable attention in recent publications, signifying the importance of research in efficient mobility for this new computing model. Moreover, the role of Mobile Edge Computing (MEC) as a key innovation in the realization of 5G cellular networks \cite{hu2015mobile} also reinforces the importance of a resource efficient mobility management system. In this section, we look at related topics in the research into mobility management for fog address both the physical handover of devices from one fog node to the next as well as the migration of services between them. 

The authors of \cite{bao2017follow} propose a framework called Follow Me Fog which introduces a handover controller in mobile devices. The controller monitors the signal strength levels observed from different fog nodes. If signal strength breaches a threshold or another fog node with better signal strength is observed, a Migration Controller in the fog node is notified. The controller is then responsible for migrating jobs from the current fog node to the target fog node before disconnecting the connection. This helps them achieve a 36.5\% reduction in latency in experimental scenarios. However, for optimal performance, this framework requires precise optimization of the handover parameters, which would be further complicated in a heterogeneous access network environment. 

The authors of \cite{bi2018mobility} propose a more all-encompassing solution that introduces an SDN layer between the cloud and fog layers. They implement signaling operations through the SDN that allows them to implement both proactive and reactive handovers. Their implementation supports proactive handovers albeit an overlap in the coverage areas of the current and future serving fog servers is required for simultaneous measurements of both fog nodes. The SDN also buffers downlink data during the handover procedures after layer 2 detachment and forwards it to the future fog server. This yields a much shorter handover latency in their simulated experiments.

A lot of research for MEC mobility also pertains to optimizing the transfer of a user's virtual machines (VM). These approaches address the trade-off between the migration cost and improvement in the quality of service (QoS) a user would observe from changing to a closer MEC server \cite{ai2018edge}. \cite{sun2017avaptive}  is one such approach that achieves this through a Profit Maximization Avatar Placement (PRIMAL) algorithm. The algorithm phrases the problem as an optimization problem solved through Mixed-Integer Quadratic Programming and places the user's Avatars in the optimal servers to reduce the handover latency.

There is also related research which deals with the use of machine learning techniques to improve the wireless handover procedure by taking a more proactive, predictive approach. \cite{tung2017big} approach the problem of forecasting future handovers and handover anomalies in future ultra-dense heterogenous 5G networks. They propose the use of machine learning based algorithms to further empower the self-organizing network (SON) system in 5G. They use a variety of popular linear and nonlinear machine learning models to forecast future handover (HO) demand including regression based models, feed-forward neural networks, and Gaussian process regression, with future predictions based on previous values observed in the time series. Their algorithms are trained to produce hourly predictions for the number of handovers to be expected for a cell. This can allow the self-organized network to adapt tuning parameters to improve or retain its quality of service (QoS) preemptively. They identify Gaussian Process Regression (GP) and Neural Network (NN) as the best models for predicting future handovers with mean absolute error values of 0.015 and 0.02 respectively.  

The authors of \cite{feltrin2018machine} show a similar method using a feed-forward neural network to learn and monitor the RSSI pattern over time. The network is trained to relate an RSSI pattern from a device with the probability that a handover going to be performed in a future time slot. They simulate a user's movement patterns using synthetically generated acceleration and speed parameters to create a dataset, with the RSS calculated using path loss and shadowing. 

A related application of RNNs is presented in \cite{wickramasuriya2017base} where they solve the problem of predicting the optimal network topology. The training data consists of RSS sequences from different base stations before a handover and the target is the optimal base station to handoff to. They test their model on an 8 class (8 base stations) problem and achieve a 98.3\% accuracy in predicting the optimal base station. 

\subsection{Our Contribution}

Several studies discuss the viability of fog computing in vehicular networks \cite{fogveh1} \cite{fogveh2} \cite{fogveh3}. However, none of these cater to accommodating the nuances fast-moving vehicles in a fog computing environment. Studies such as  \cite{bao2017follow}, \cite{bi2018mobility} and \cite{wickramasuriya2017base} also test their approaches using synthetic user movements which simulate users and vehicles moving within tight parameters. In contrast to this, our test simulations utilize a dataset of actual vehicle movements. This allows us to gauge how the learning models perform in urban and suburban vehicular movement. Moreover, our proposed solution also accommodates mobile fog nodes, as the fog location is taken into account by the learning models in the training data. 

Our use of the cost metric, the latency for a service request, at the application layer also makes it more suitable for deployment over heterogeneous networks. This metric accommodates the effect of the variability of wireless channels on performance into the handover scheme which can be a major problem for latency driven tasks \cite{mao2017survey}. This is in contrast to proposed approaches such as  \cite{feltrin2018machine} \cite{bao2017follow} that focus on RSSI and threshold-based handovers, which would have to be adapted and optimized based on the signal strength of the radio access technology. Our request routing scheme also does not require a coverage overlap for a preemptive handover as required in both \cite{bao2017follow} and \cite{bi2018mobility}. As pointed out in \cite{tung2017big}, a system such as ours would be invaluable in providing seamless connectivity for network management in conjunction with a SON mechanism.  The machine learning models we designed can also use data from several days over multiple hours, which allows our system to learn the variation in performance over the course of the day. This is a highly important feature as usage patterns, and therefore server loads, vary throughout the day as well as the week \cite{tung2017big}.

\section{System Architecture} \label{architecture}
The machine learning based handover optimization system presented here is envisioned as a part of JAMScript \cite{wenger2016programming}. The JAMScript project is a polyglot programming language for IoT computing that caters to the hierarchical and distributed architecture of edge computing. The language follows a three-layer hierarchical structure in keeping with the edge computing architecture including the cloud, fogs and edge devices. Moreover, the programs follow an SDN-like controller-worker model with a program overlapping all three entities, allowing for easier interoperability and control in distributed applications. 

Amongst its many features, JAMScript offers the notion of adjacency between computing nodes, similar to that in aggregate computing \cite{beal2015aggregate}. Such a programming paradigm is highly suited to the IoT scenario where the control and collaboration of numerous devices can be invaluable but quite challenging. It allows for nodes that are adjacent to each other to readily share data and processing tasks. This aids the easy collaboration of devices on tasks, actualized through the removal of task invocations and transfer of data between nodes. The JAMScript model works on a hierarchical adjacency in keeping with the edge computing architecture. In this model,  fog nodes are adjacent to the cloud and edge nodes are adjacent to fogs. 

Unlike cloud computing with its centralized resources, fog computing brings resources closer to the edge devices via fogs deployed in a distributed fashion. As a consequence, certain fog nodes would be closer to an edge device than others. Therefore, the adjacency is an important feature to support mobile device in an edge computing scenario. Especially for highly mobile devices with variable motion such as vehicles operating in traffic, the fog associations can change in a rapid and highly variable manner. This makes handover optimization an important requirement to minimize the disconnection time as a device disconnects from the current fog node and reconnects with the next fog node. 

Such service disruptions can have a marked impact on both the device and the fog node's performance. For devices, it could limit access to important tasks offloaded to the fog nodes and potentially limit their functionality. For fog nodes, the impact would be variable depending on the mapping of processing tasks. For instance, for a distributed application where the fog is computing results based on computations from multiple edge devices as inputs, disconnections can limit these inputs and thereby affect the quality of the results. To truly maximize the potential of fog nodes, being able to predict disconnections, reconnections and the relative costs of different fog nodes can be invaluable. Having such information reliably available can allow devices to select the optimal fog node at any given location and time, defer task invocations and to offload important task executions in advance. 

\begin{figure}
\includegraphics[width=\linewidth]{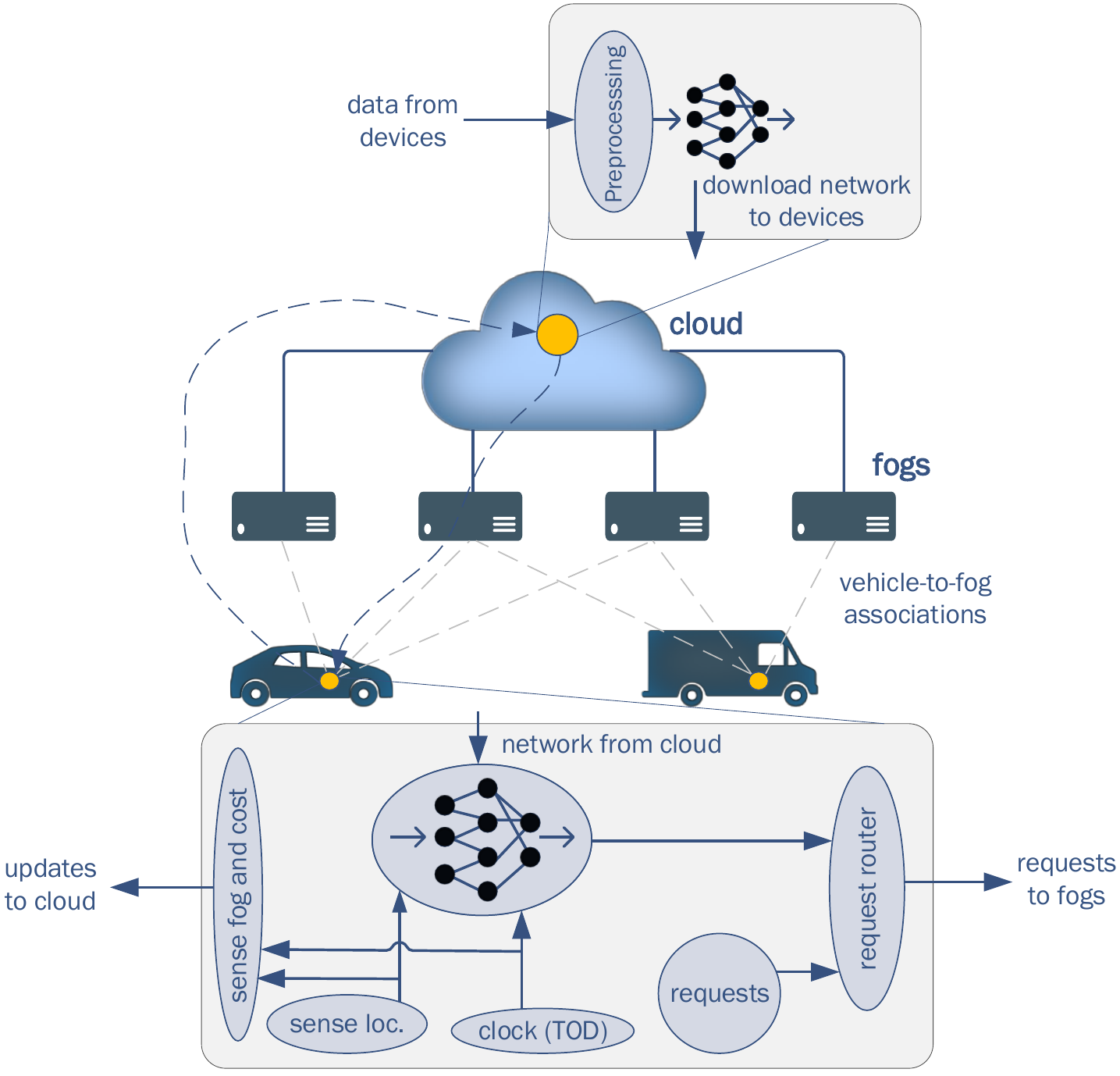}
\caption{A system diagram showing the flow of information of our system through an edge computing architecture. }
\label{fig:systemdiagram}
\end{figure}

Figure \ref{fig:systemdiagram} shows the overall architecture of our system and how it can provide such predictive information to the devices to empower their decisions. Our system takes a data-driven approach to model the ideal fog association and its corresponding cost at a given location and time. This requires sensing and transmitting these feature for a given location and time to the cloud where it can be used to train the neural networks to create a pair of continually learning models. These features can be generally readily available as feedback for a vehicle's interaction with a fog node. The models can be periodically downloaded to the edge devices to limit their interaction with the cloud and fogs and allow them to make decisions locally. Having preemptive information about the ideal fog node at a given time and location can radically improve handover performance and allow the device to optimize the routing of requests to the fog nodes to maximize their potential.

Our system is designed to work in conjunction with an existing handover mechanism and provide predictive services to allow proactive handover decisions. For example, for our simulations, we use the distance of a vehicle from the fog nodes as a handover criteria as described in section \ref{subsec:experimental}. Our learning models learn the handover behavior through fog associations based on this underlying handover policy as vehicles navigate through the fog network. Therefore, failure of fog nodes and handling of newly commissioned fog nodes would still be handled by the underlying handover mechanism.

\section{Fog Location and Cost Prediction} \label{sec:fogandcost}

At the core of our design are two neural network based models trained to learn the fog node and the cost of interacting with it at a given location and time. Both of these models are trained using data sensed from device interactions with fog nodes. In this section, we give an overview of the models.

\subsection{Fog Prediction Model}

We treated the fog prediction problem as a multi-class classification problem where task execution times for devices at a given time and location can be associated with a finite set of possible fog nodes. 

A preliminary step before the learning process is phrasing the raw data into a supervised learning problem. Supervised learning is a common class of machine learning methods where the model is provided an input and output pair and the model learns the association between them. For our scenario, inputs for the fog predictor were the coordinates of the device, the coordinates of the fog node, the time of day of the interaction. The output was the label of the fog node the device had been interacting with. 

A multitude of machine learning algorithms have been shown to be able to solve this type of problem \cite{kotsiantis2007supervised}. We chose to use a feed-forward neural network as a multi-class classifier for our application. Multilayer neural networks are universal function approximators  \cite{hornik1989multilayer}, capable of learning complex behaviors and even surpassing human-level performance \cite{he2015delving}. They also tend to thrive in data-rich applications such as ours where we can have millions of device interactions in a smart city scenario. 

We compared the performance of our neural network on our test dataset with two other popular classification algorithms: Support vector machine (SVM) and K-nearest neighbors (KNN). All of the algorithms tended to perform well in terms of classification accuracy with the neural network showing an edge in prediction accuracy. Table \ref{tab:fogcomparison} gives a summary of results. In addition to the performance advantage, neural networks offer the added benefit of being quite portable and scalable, making it easier and less expensive to transfer copies of a trained network to devices for local use. In contrast to this, instance-based learning algorithms such as KNN require all the data points to make an accurate decision, thereby requiring the entire dataset to be available locally.

\begin{table}
  \caption{Fog predictor: comparison with other learning algorithms.}
  \label{tab:fogcomparison}
  \begin{tabular}{ccl}
    \toprule
    Algorithm & Percentage accuracy \\
    \midrule
    Feed-forward neural network & 99.2\%  \\
    KNN classifier with 4 nearest neighbors & 98.42\% \\
    SVM classifier with RBF Kernel & 98.1\%  \\
  \bottomrule
\end{tabular}
\end{table}

For our fog prediction model, we used a neural network with three hidden layers. Each hidden layer contains 100 neurons all using the sigmoid activation function. The output layer has neurons representing each class or fog node. The details of how this architecture was chosen are shared in Section \ref{sec:fogimplementation} and the detailed results are shown in Figure \ref{fig:confusionmatrix} as a confusion matrix.

Feed-forward neural networks offered us a lot of competitive advantages for our application. One of the biggest advantages was the ease of adapting them into a continual learning system that takes into account fog nodes entering and leaving the system. As new fog nodes enter the system and old ones leave, the neural network can be selectively retrained and expanded with well-documented approaches such as \cite{lee2017lifelong}.

\subsection{Cost Prediction Model}

Modeling the cost for the devices' interactions with the fog node required a different approach than the one for the fog predictor. This was partly down to the cost being a real value with an infinite number of possible values as well as to cost modeling being a more complex problem that required a more capable learning model. 

\begin{figure}
\includegraphics[width=\linewidth]{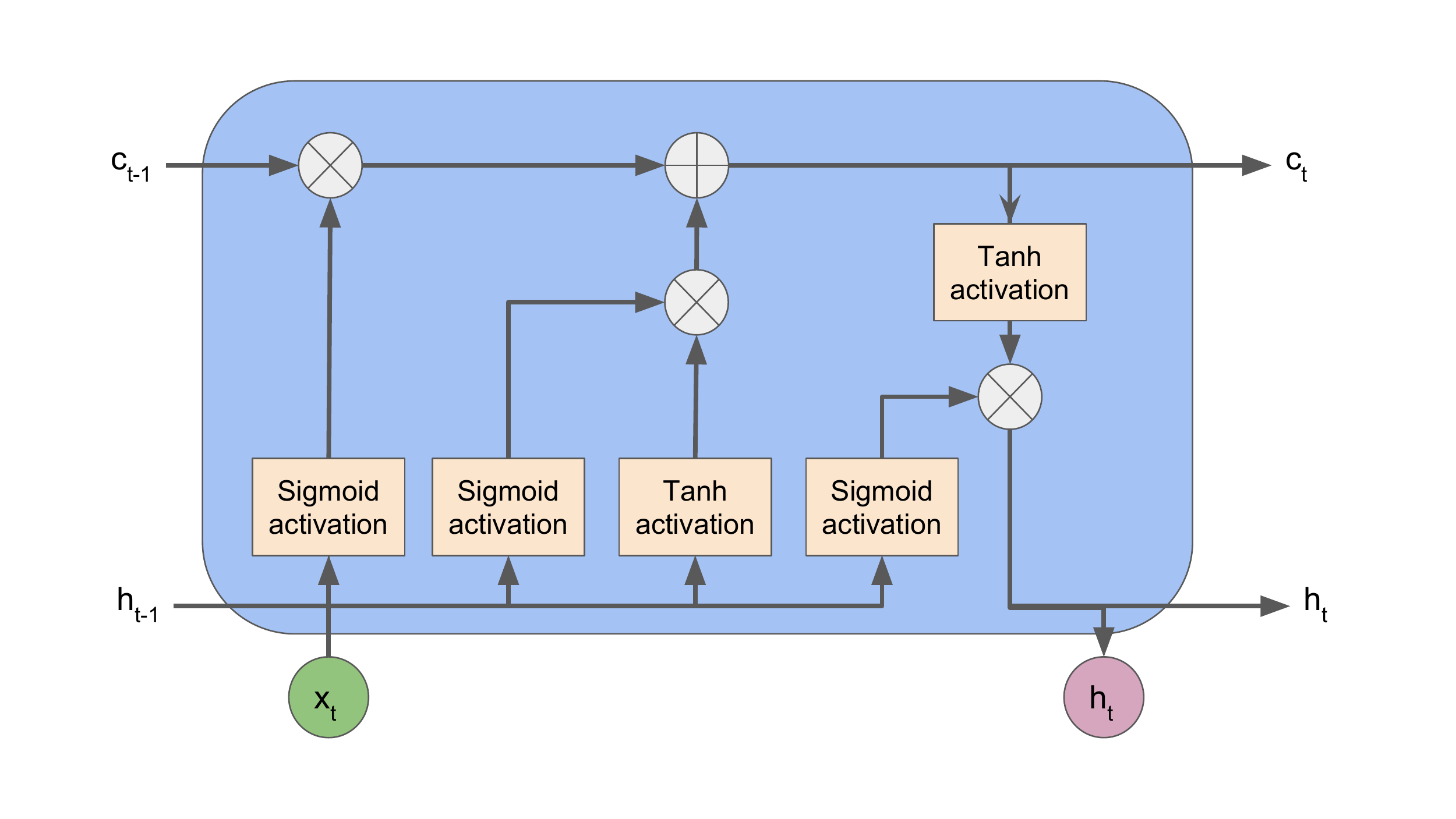}
\caption{A long short-term memory cell. In addition to the input variable $x_t$, the hidden state of the previous LSTM cell, $c_{t-1} $ as well as its predicted output, $h_{t-1} $are all taken into account in mapping the input to the output. Such cells can be cascaded to learn associations in series data}
\label{fig:lstm}
\end{figure}

To suit these requirements, the prediction model is implemented as a two-layer stacked recurrent neural network with long short-term memory cells \cite{hochreiter1997long}. RNNs are a variant of neural networks that are expanded in time, which allows them to pick up temporal patterns in time-series data \cite{connor1994recurrent}. However, they are also more susceptible to the vanishing gradient problem where the influence of older values in the series diminishing and limiting their ability to learn long-term dependencies. By replacing the neurons with LSTM cells like the one shown in Figure \ref{fig:lstm}, this problem can be overcome and allow such networks to learn long-term dependencies in the data more effectively \cite{greff2017lstm}.  

The different model also required rephrasing the supervised learning problem to suit the architecture of the RNN. Each training sample consisted of a series of inputs from the device with the future cost value as a target. The cost at a future point was therefore learned as a function of the device's location, fog location and interaction cost of the last 10 data points. 
The choice of 10 previous values was determined using the autocorrelation and partial autocorrelation plots of the data that show the measure of the correlation of a variable with its previous values and verified through cross-validation. 

The architecture of the RNN reflected the shape of the input, with each of the stacks containing 10 LSTM cells. The output from the RNN was fed into a feed-forward neural network with just one hidden layer containing 20 neurons with a sigmoid activation function. The output layer contained a single neuron using a linear activation function to accommodate the real-valued cost. 

A comparison of the performance of the RNN with other learning algorithms in Table \ref{tab:costcomparison} shows the performance advantage this approach offered. Section \ref{sec:testresults} gives a more detailed overview of the performance of the cost predictor.

\begin{table}
  \caption{Cost predictor: comparison with other learning algorithms.}
  \label{tab:costcomparison}
  \begin{tabular}{ccl}
    \toprule
    Algorithm & Mean absolute error \\
    \midrule
    Feed-forward neural network & 0.0518 \\
    RNN with LSTM cells & 0.0346 \\
    KNN with 6 nearest neighbors & 0.0517 \\
    SVR with RBF Kernel & 0.0590 \\
  \bottomrule
\end{tabular}
\end{table}

\section{Request Routing} \label{sec:requestrouting}

Using our two models, we propose a request routing system that would allow the seamless transition of a mobile device from one fog node to the next with minimal interruptions. 

A prerequisite for this system is knowing the trajectory of the device in advance. This is a reasonable expectation given the common use of navigation software in vehicles. 

This approach requires feeding in a device's trajectory point-wise into the trained fog predictor to know the identity of the serving fog node at each point in its motion. As copies of the learning models are available locally to the device and periodically updated, this task can be performed locally in conjunction with the associated fog nodes independent from the cloud. 

\begin{figure}
\includegraphics[width=\linewidth]{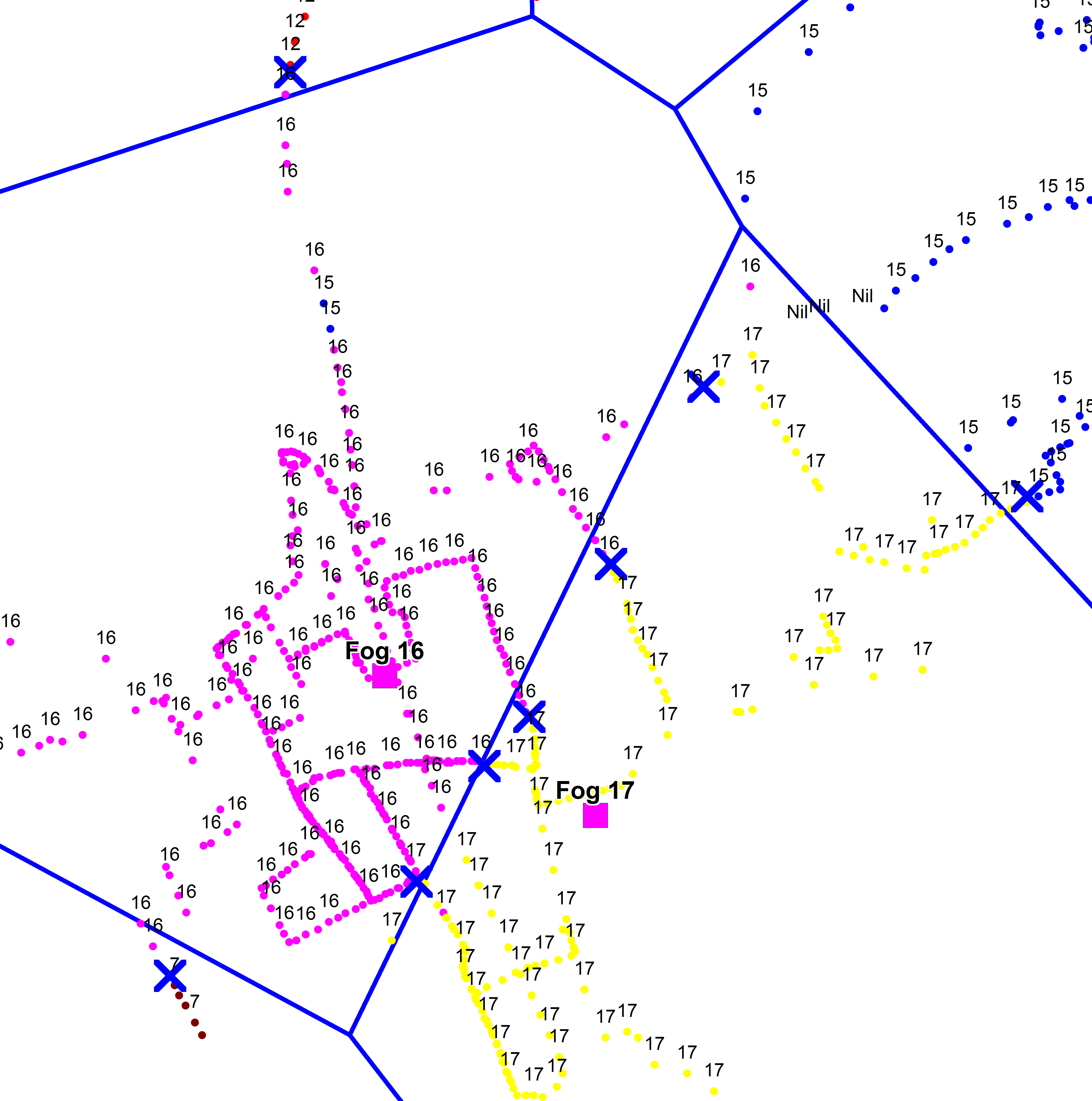}
\caption{Transition points as predicted by the fog predictor. The transition points are represented by blue crosses. The lines identify the cell boundaries for the fog node's serving area. }
\label{fig:transition}
\end{figure}

Figure \ref{fig:transition} shows the performance of the prediction algorithms in predicting a transition from one fog node to another. To accomplish this, we fed a test device's movements into the predictor and located the point where the predictor's output transitioned from one fog node to the other. We used a heuristic approach to identifying the transition point along a device's trajectory by observing the predicted fog associated with the previous three and future three data points. A transition from one fog node to another in these points identifies the location as a transition point. As evident in Figure \ref{fig:transition}, the handover prediction mechanism correctly identifies all the transition points along the cell boundary for the fog nodes. 

To accommodate lower sampling rates for data sensing and ambiguity in predicting the correct fog node along the cell boundary, we introduce the concept of a buffer zone before the predicted handover point. When the buffer zone is breached, current fog node can be notified by the device and the transfer of offloaded task and data can be initiated to the future fog node. Similarly, the device would start routing requests to both the current and future fog nodes to minimize the service interruptions.

\section{Implementation}\label{sec:implementation}

A proof of concept implementation was done in Python for this project. The popular Keras \cite{chollet2015keras} API was used to create the deep learning models using a Tensorflow backend. Scikit-learn \cite{scikit-learn} was used for implementing the other machine learning algorithms for testing and comparison purposes. The details of the implementation are as follows:

\subsection{Input Matrix Design and Preprocessing}

In addition to phrasing the raw data into a supervised learning problem to train the neural network, a variety of preprocessing techniques were applied to ease the learning process. From each fog service request, the following features are extracted: coordinates of the device, the interacting fog node and its coordinates, the time of day and cost of the service request.

The training target for each model is different since they have to learn different behaviors. For the fog prediction, we use one hot encoding to identify the fog node being interacted with. For the cost prediction model, the scaled cost value is used as a target.

In the preprocessing phase, the distance of the fog node from the device is extracted. In our actual simulation, this was provided by the simulator. The geographical coordinates for both the fog and device are broken down into Cartesian x, y and z components as we observed this made it easier to learn patterns for the models. The time is also broken down to the day of the week and time of day in seconds as traffic patterns tend to vary with these parameters. 

A further feature mapping is applied to the time, where it is represented by its sine and cosine components rather than the ordinal value in seconds. This allows the cyclic nature of time to be incorporated into the learning process. Lastly, all the real-valued input data, as well as the target cost, was scaled between 0 and 1.

\subsection{Fog Predictor} \label{sec:fogimplementation}
The fog predictor was implemented as a three layer feed-forward neural network. Each of the three hidden layers contained 100 neurons and the ideal network architecture was identified using cross-validation. 

The Adam optimizer \cite{kingma2014adam} was used along with the categorical cross entropy loss to train the model. The model was trained for 3000 epochs till no accuracy gain was observed on the validation set. In general, the model showed no overfitting over the course of training and using a dropout layer for regularization \cite{srivastava2014dropout} produced no gains in performance even with prolonged training.

\subsection{Cost predictor}

The more complex nature of the RNN architecture required a different implementation process. Sequential input data from each vehicle was converted into an input matrix. Each sample, or row in the input matrix, consisted of the 10 input vectors associated with each of the 10 last data points in the vehicle's trajectory. The target for the row was the cost value for the future data point. For the start and end points of each device's trace, zero padding was used to fill to accommodate the lack of values in the row. 

As with the fog predictor, the Adam optimizer was used to train the model. The mean absolute error was used as the loss function and yielded the best results. 

\section{Experiments and Analysis} \label{sec:exp}

To evaluate the capabilities of the prediction models, we carried out a set of experiments. All these experiments were based on the dataset derived from the JAMScript emulator.

\subsection {Creating the Dataset and Simulating the Network} \label{subsec:experimental}
The datasets for training and testing the machine learning models were generated using taxi traces from Shanghai Jiao Tong University \cite{dataset}.  In this trace, actual taxis were instrumented to obtain their locations at predefined time intervals throughout a day. For occupied taxis, reports were generated every second and for unoccupied taxis, they were generated every 15 seconds. However, the dataset consisted of just movement traces with no fog network associated with them. 

A vehicular fog system simulator written in JAMScript was fed these traces to obtain data for training and testing our machine learning models. The vehicles and fogs are represented in the simulator using Docker containers. The inter-container access latency is set using the Linux NetEm modules at each container according to the distance between the fog and the vehicle. This way we can accurately model the latency of access from the vehicles for requests reaching the different fogs. The inter-container latencies are continuously updated to reflect the changing locations of the vehicles with respect to the fogs. 

In the simulation, we place fogs in selected points in the city map and drive the cars according to the trace. Using fog selection capabilities in JAMScript, the simulator keeps the vehicle associated with the nearest fog as it moves according to the trace. Services are handed over from a current fog node to the future one when the vehicle's distance from the future fog node is smaller compared to the vehicle's distance from the current fog node. The simulator measures the response times the vehicle observed for executing remote tasks in the assigned fog. In certain locations, the vehicle could find no fogs or could be transitioning between fogs. The fog locations were chosen to simulate a real-world deployment with a larger number of fogs in the dense urban areas and fewer in the suburban areas. We ran the simulation for 53 vehicles and 20 fog servers, yielding 230,000 data points which gave us a healthy dataset for training and testing our models.

\begin{figure}
\includegraphics[width=\columnwidth]{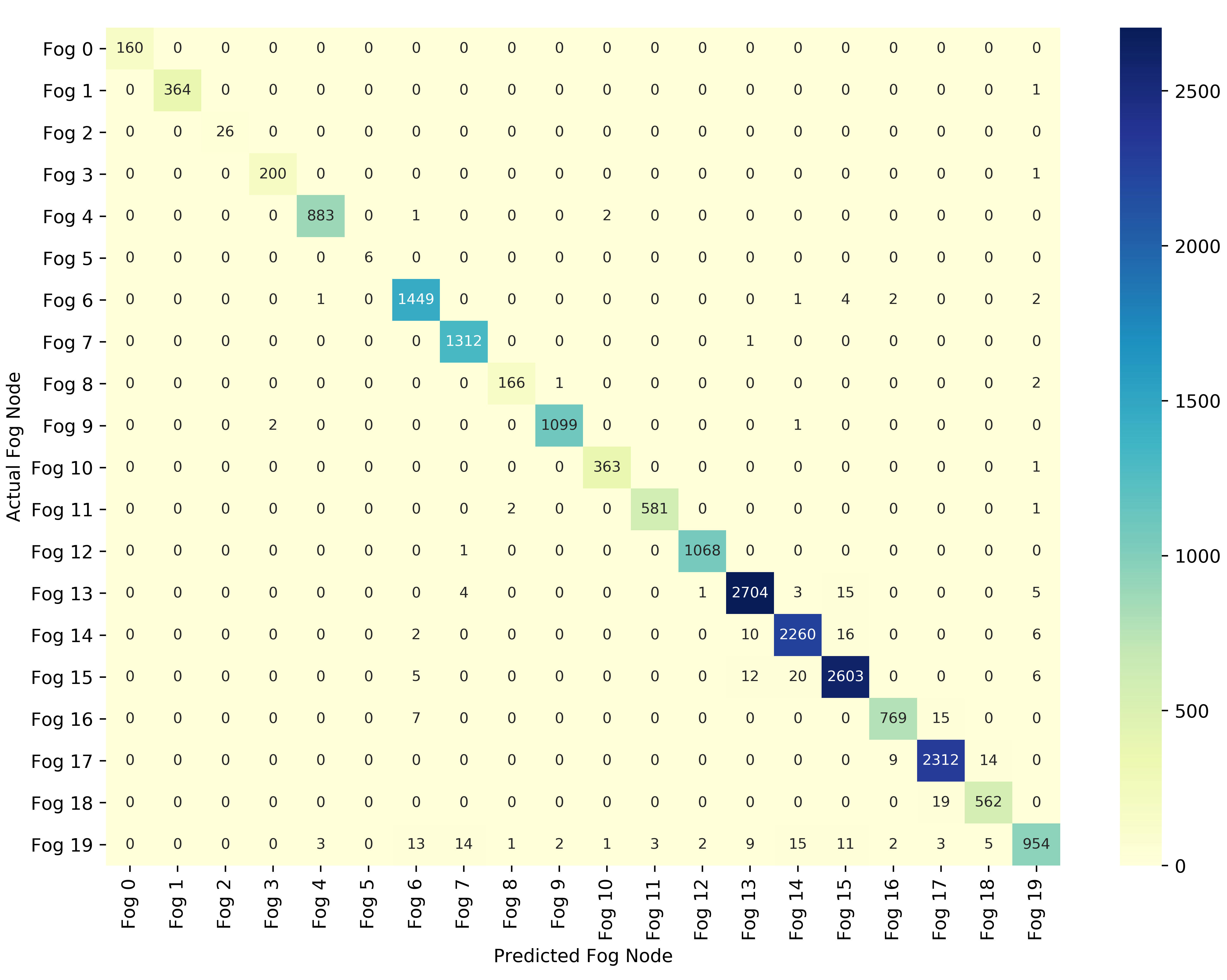}
\caption{Confusion matrix for the fog classifier's performance on a test set. The y-axis represents the actual fog node label and the x-axis represents the predicted fog node label. Quantities lying along the diagonal were correctly classified and quantities lying off-diagonal were incorrectly classified. The map also reflects the class distribution of data points for different fog nodes.}
\label{fig:confusionmatrix}
\end{figure}

\subsection{Performance on a Test Set} \label{sec:testresults}
As is standard in machine learning, we divided our data into training, validation and testing sets. The test set for the fog predictor consisted of 20,000 data points while the test set for the cost predictor consisted of 30,000 data points to aid visualization. 

The fog predictor was able to accurately predict the correct fog node for the test data with an accuracy of 99.2\%. Figure \ref{fig:confusionmatrix} shows the confusion matrix for the model's performance on the test set. Our dataset had a balance of high and low usage fog nodes as is typical of a real-world scenario. 

\begin{figure}
\includegraphics[width=0.8\linewidth]{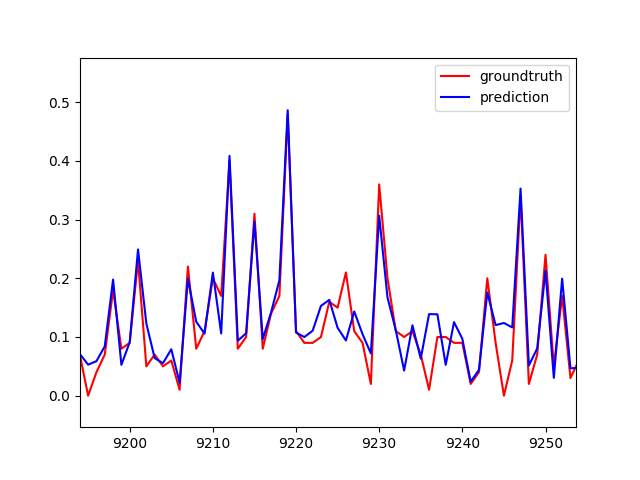}
\caption{Graph of the predicted value vs the actual value for the cost from the test dataset.}
\label{fig:costpredgraph}
\end{figure}

\begin{figure}
\subfloat[Actual cost of the test.]{%
  \includegraphics[clip,width=\columnwidth]{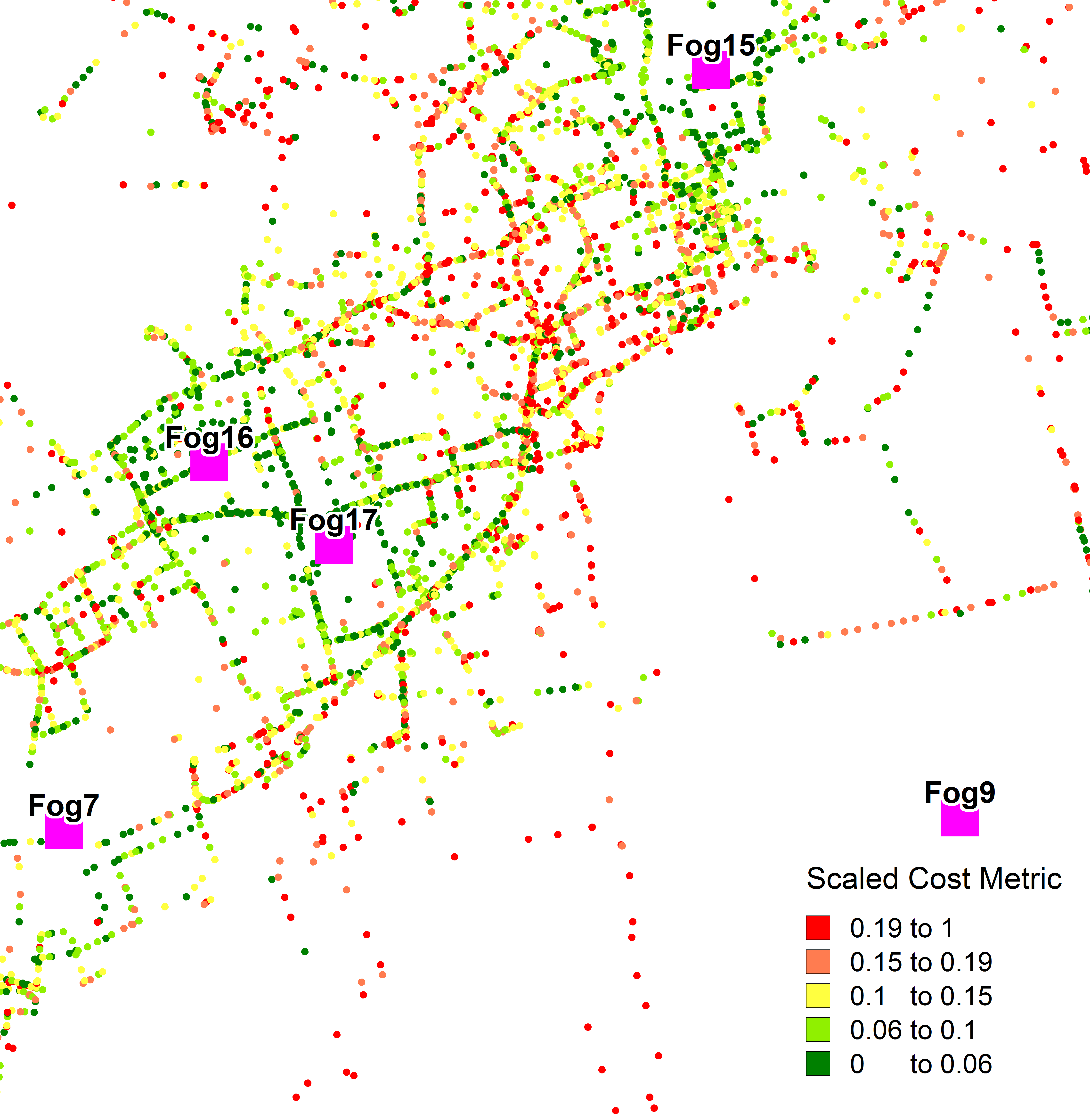}%
  \label{fig:actualcost}
}

\subfloat[Predicted cost from the cost model on the test set.]{%
  \includegraphics[clip,width=\columnwidth]{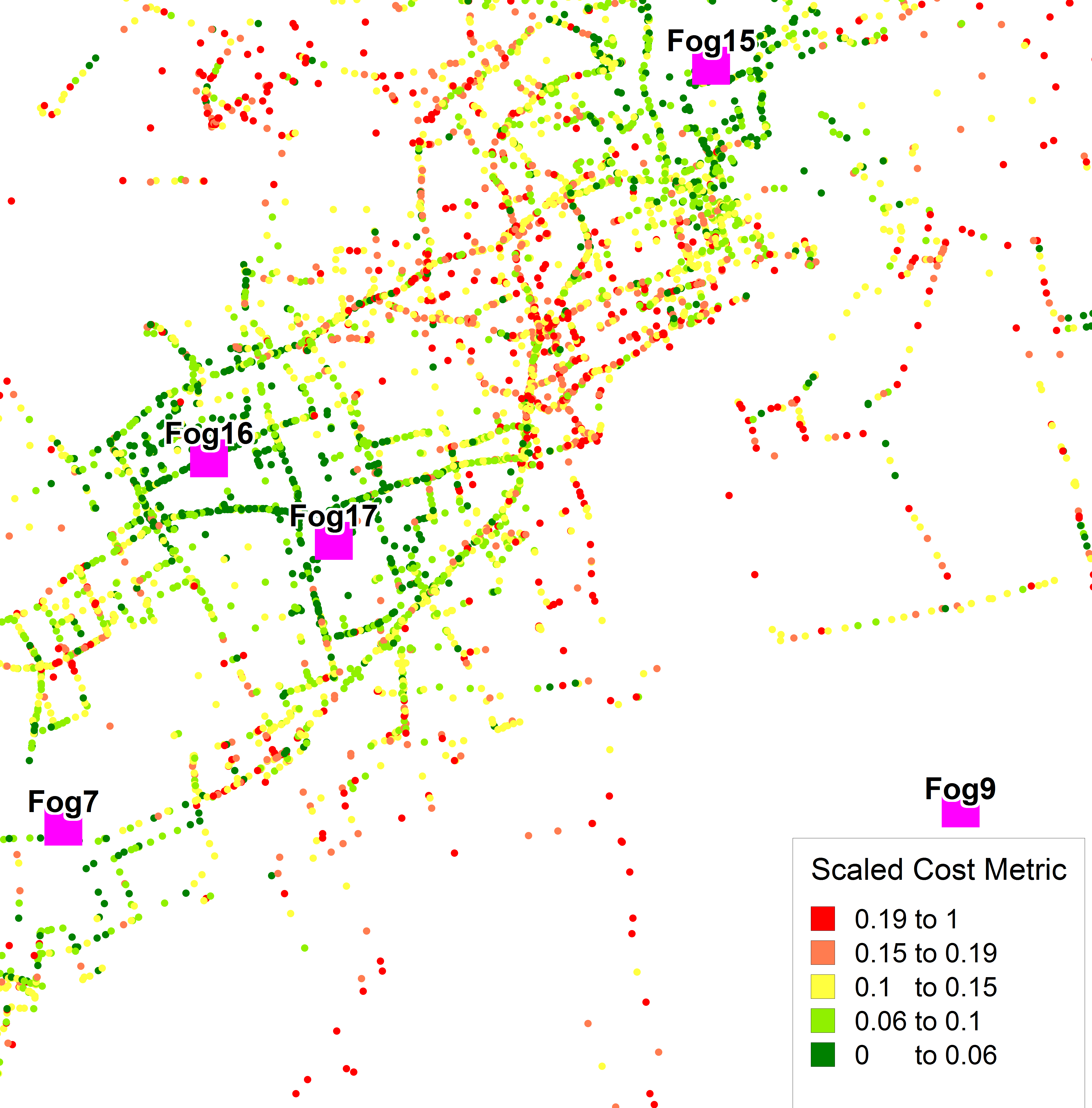}%
  \label{fig:predictedcost}
}
\caption{A comparison of the actual cost with the prediction on the test set. Fog nodes are highlighted as magenta blocks. The model picks up on the general trend of the cost increasing as vehicles move further away from the fog nodes, as well as any clusters with low performance along the cell boundaries. }

\end{figure}

The cost predictor was proficient in picking up the dominant trends in the training data. Figure \ref{fig:predictedcost} shows a plot of the predicted cost metric for the test set in comparison to the actual cost values in Figure \ref{fig:actualcost}. Data points in the close vicinity of the simulated fog nodes show a low cost. The cost value tends to increase as the data points move away from the fog nodes and towards their cell border. Figure \ref{fig:costpredgraph} shows the prediction tracking the ground truth value for a section of the test dataset.

\subsection{Urban and Suburban Segregation}

We extended our analysis to how the learning models performed in the different demographics offered by our dataset. To keep the experiment accurate, we had more fog nodes deployed close together in the urban city center and fewer ones spaced further apart in the suburban areas. The results for the segregated test set for the fog and cost predictors are shown in Tables \ref{tab:fogurban} and \ref{tab:costurban} respectively. 

\begin{table}
  \caption{Fog predictor performance: urban and suburban segregation.}
  \label{tab:fogurban}
  \begin{tabular}{ccl}
    \toprule
    Demographic& Percentage accuracy \\
    \midrule
    Urban & 98.34\% \\
    Suburban & 99.65\% \\
    Cumulative & 99.2\% \\
  \bottomrule
\end{tabular}
\end{table}

\begin{table}
  \caption{Cost predictor performance: urban and suburban segregation.}
  \label{tab:costurban}
  \begin{tabular}{ccl}
    \toprule
    Demographic& Mean absolute error in Cost \\
    \midrule
    Urban & 0.028 \\
    Suburban & 0.041 \\
    Cumulative & 0.034\\
  \bottomrule
\end{tabular}
\end{table}

\begin{figure}
\includegraphics[width=0.8\linewidth]{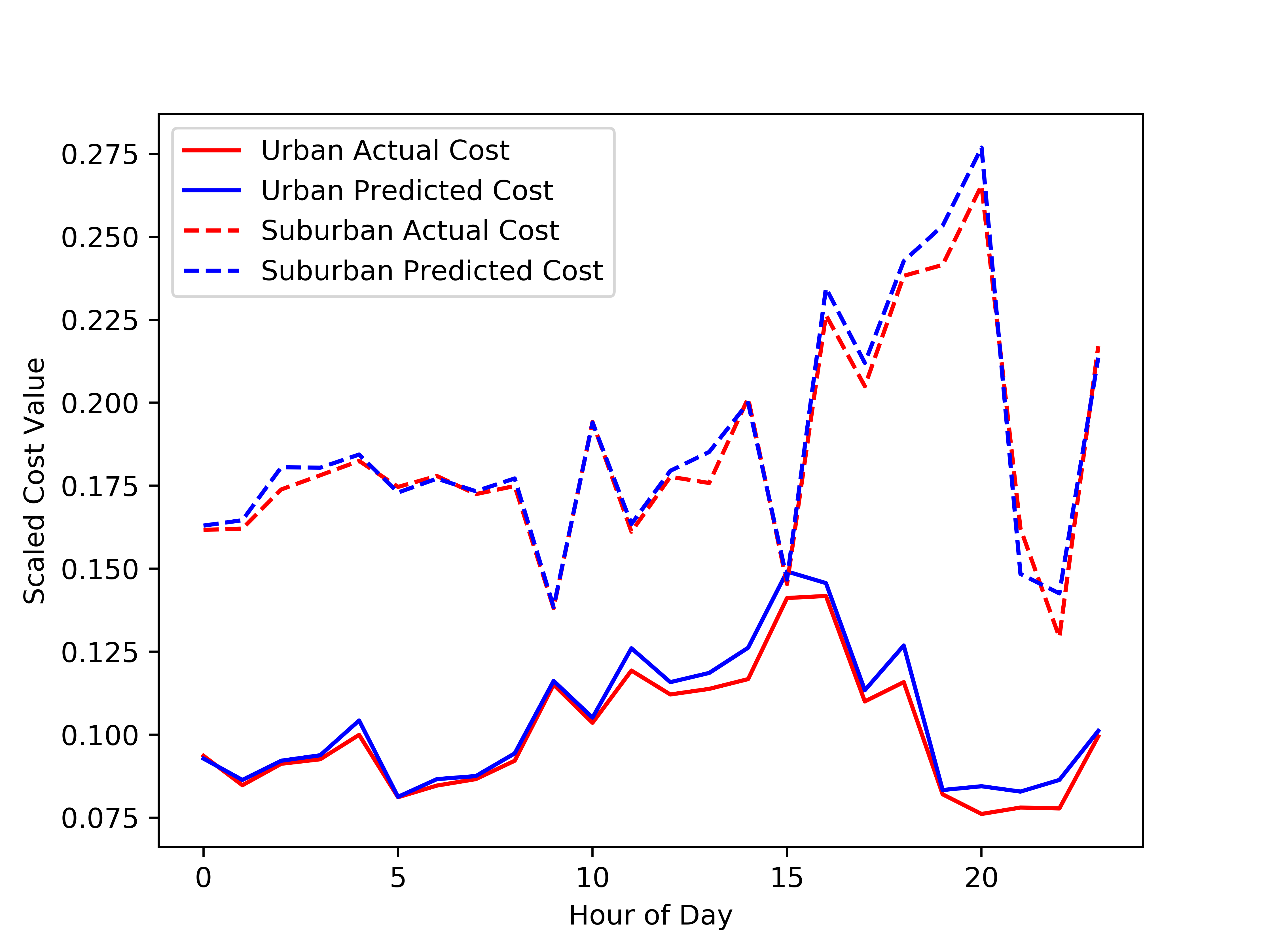}
\caption{Graph of the average predicted value vs the average actual value for the cost on the test set as a function of time. The urban and suburban demographics have been segregated. Over the course of the simulation, the cost tends to fluctuate due to loading and movement of the cars. Our model picks up these temporal trends as well.}
\label{fig:temporalgraph}
\end{figure}

We noticed a higher accuracy for the fog predictor in the suburban areas than in the urban areas. We reasoned this down to having more fog nodes in close vicinity creates smaller cells with more boundaries in an urban setting. Misclassification by the model is more probable along the boundaries, albeit by a small margin. For the cost predictor, performance gauged by the mean absolute error between the predicted and actual output was better in the urban areas. This can be reasoned due to the more consistent cost with less variation being easier to track for the learning model in contrast to the suburban setting where the cost tended to have a much higher variance. This is verified in the following section as well as the cost plot for urban and suburban traffic in Figure \ref{fig:temporalgraph}.

\subsection{Learning Temporal Patterns}

Cost values are susceptible to fluctuation for any location due to a variety of factors such as the number of devices submitting service requests. We gauged our model's ability to learn temporal variations in the cost value over the course of the day. Figure \ref{fig:temporalgraph} shows the results for our analysis. The model appears to be able to follow the variation in cost with respect to time for both urban and suburban areas in our test set. The error rate tends to increase because training data decreases as the simulation progresses, as more of the taxis complete their traces, resulting in fewer training data points for the latter hours compared to the earlier hours. 

\subsection{Obstacle Detection}

One of the aims of our system was to be able to learn any obstacles or low coverage areas. Knowing low or null coverage areas such as tunnels in advance could be an invaluable feature for a resource management system as data transactions can be planned in advance to avoid data loss. 

\begin{figure}

\subfloat[Simulated tunnel added into the training data.]{%
  \includegraphics[clip,width=0.8\columnwidth]{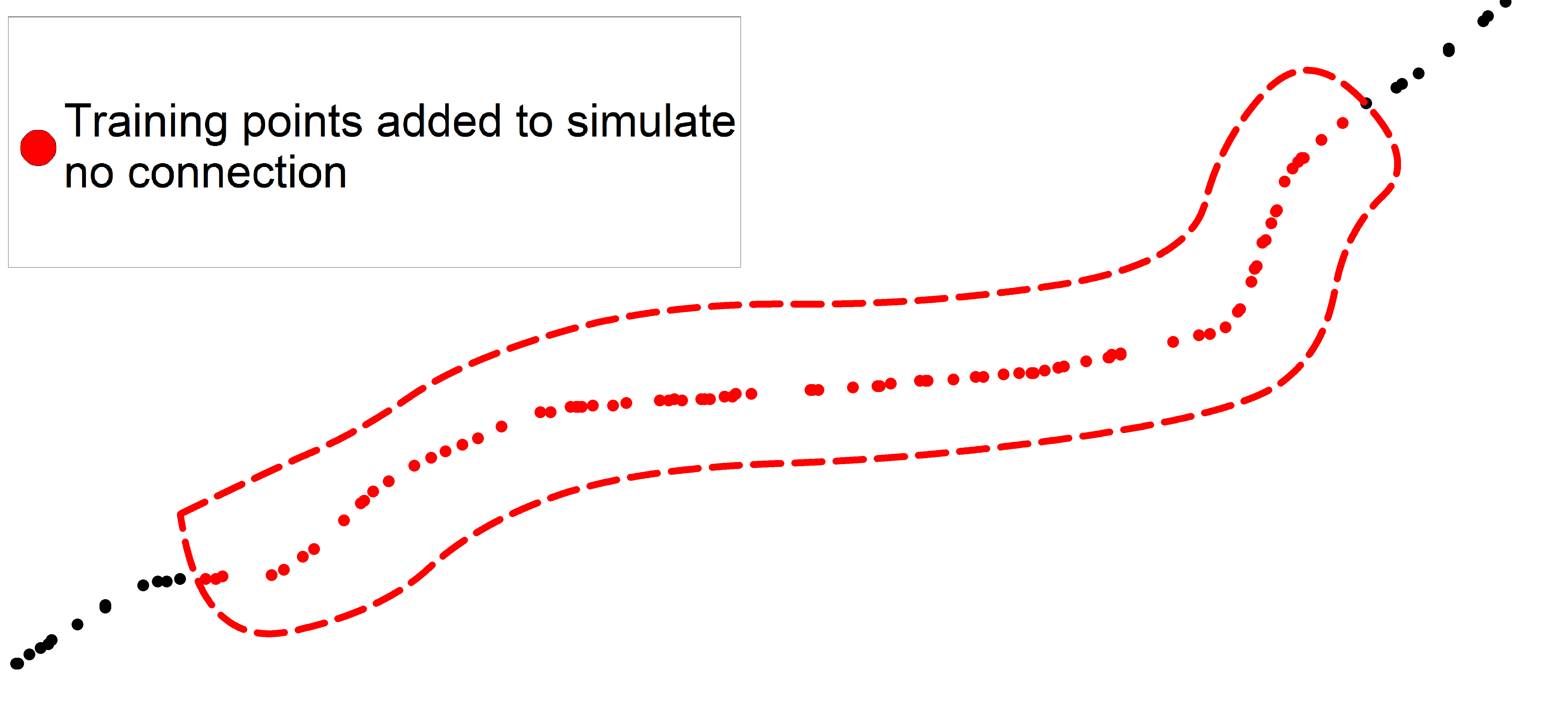}%
  \label{fig:trainingnull}
}

\subfloat[Test points along the simulated tunnel.]{%
  \includegraphics[clip,width=0.8\columnwidth]{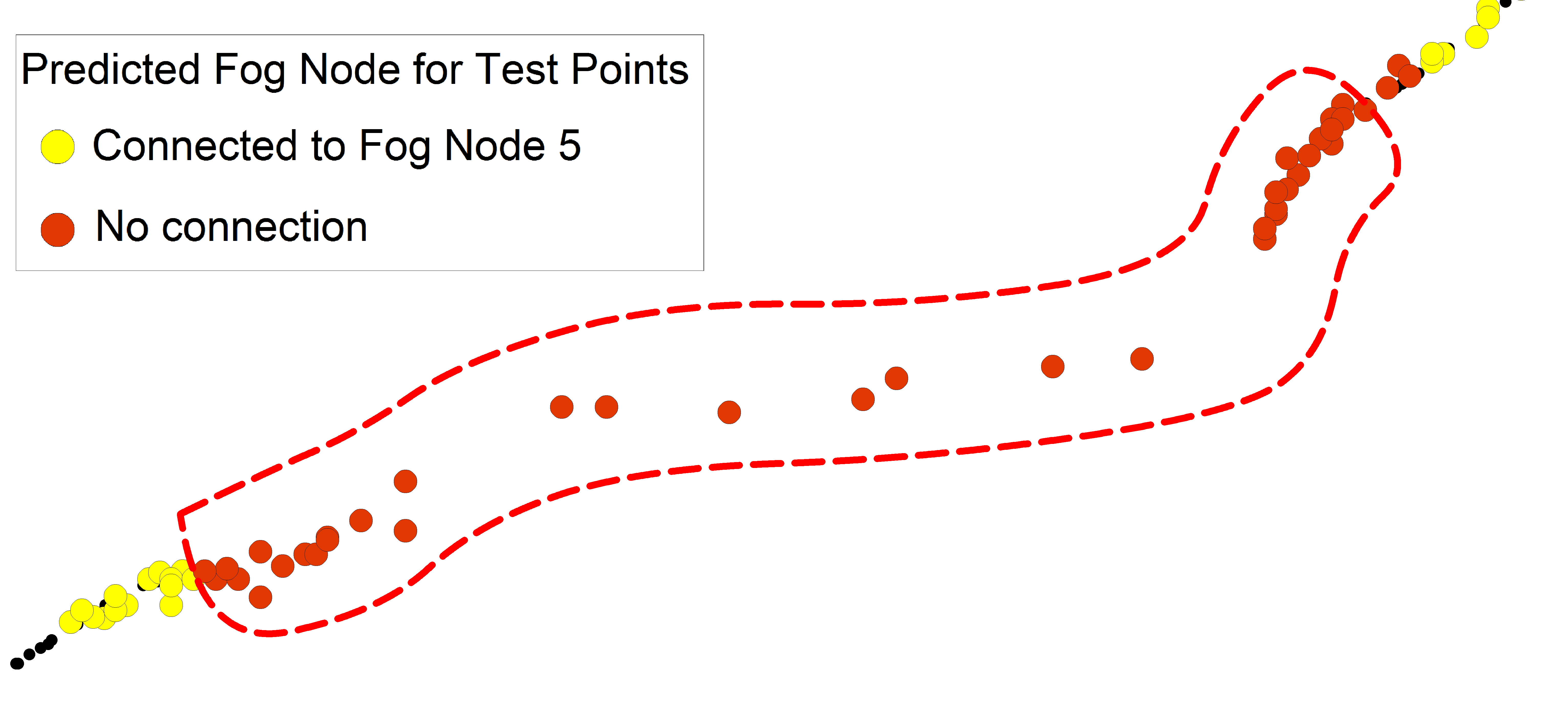}%
  \label{fig:testingnull}
}

\caption{Training the network on a low coverage area and testing the performance. Orange points represent data points classified as no coverage or infinite cost. The yellow points represent an active fog node serving the device at the location. }

\end{figure}

We produced synthetic patches of no coverage represented by an infinite cost and injected them into the training sequences for the feed-forward neural network. Figure \ref{fig:trainingnull} shows one such patch that was injected into the training data for the fog predictor. Test points were then chosen from the anomalous area and its surrounding to gauge how well the predictor had learned the boundary for the problematic area. Figure \ref{fig:testingnull} shows the performance gauged on test points. The neural network correctly identifies points along the boundary with a slight overshoot at one end.  

As the request routing system queries the fog predictor with a vehicle's trajectory, it can know in advance the location of any areas where service interruptions would be inevitable. This can allow it to preemptively offload tasks to the fog nodes serving before and after the low coverage area as well as conserve fog and device resources by lowering the frequency of repeat requests and other error correcting measures. 

\section{Conclusion and Future Work}

In this paper, we proposed the use of machine learning approaches to aid mobility management for fog computing in the Internet of Vehicles. Performance on a test set and experiments show that our system can allow a vehicle to accurately predict handover points for fog nodes as well as areas of low coverage. Moreover, our cost prediction model is also able to forecast a fog node's performance at a given location and learns both geographical and temporal patterns in the data. We further propose how such a system can be used for more efficient resource routing.  

This approach shows promise in enabling smarter resource routing schemes for fog computing. Future work could focus on learning additional features associated with service requests such as the number of vehicles working on synchronous tasks at a given location. The performance of these models in mobile fog computing scenarios, where fog servers are mounted on vehicles, also requires investigation as well. Moreover, integrating the offloaded task complexity by classifying tasks or introducing a metric for program complexity could make the cost prediction more accurate and useful for the vehicles and fog infrastructure. Our research also showed the variation in performance between different demographics. This could be exploited in future work, where regions similar regions can be clustered and trained on their personal neural networks. Such a scheme could yield better prediction performance as well as the ability to exploit transfer learning, where a trained neural network can be deployed in a new cluster. Lastly, we intend to deploy this system within JAMScript and showcase the advantages of handover optimization for different IoT applications in simulations and testing in real-world conditions.

\begin{acks}
The authors would like to thank the Natural Sciences and Engineering Research Council of Canada for supporting this work.

\end{acks}

\bibliographystyle{ACM-Reference-Format}
\bibliography{sample-bibliography}

\end{document}